\documentclass[onecolumn]{webofc}
\usepackage[varg]{txfonts}   

\usepackage[labelfont=bf, font=small]{caption}

\DeclareCaptionLabelSeparator{dot}{.\space}
\usepackage{subcaption}
\usepackage{siunitx}
\usepackage{xcolor}
\usepackage{hyperref}
\hypersetup{
    colorlinks = true,
    linkcolor  = blue,
    filecolor  = blue,
    citecolor  = blue,      
    urlcolor   = blue,
}

\newcommand{\R}[0]{\mathcal{R}}
\newcommand{\D}[0]{\mathcal{D}}

\newcommand{\N}[0]{\mathcal{N}}

\newcommand{\OO}[0]{\mathcal{O}}

\newcommand{\ww}[0]{\mathbf{w}}

\begin{document}

\title{Triggerless data acquisition pipeline for Machine Learning based statistical anomaly detection}
%\subtitle{Abbiamo un subtitle?}

\author{
    \firstname{Gaia}   \lastname{Grosso}\inst{1,2} \and
    \firstname{Nicolò} \lastname{Lai}\inst{1,2} \and
    \firstname{Matteo} \lastname{Migliorini}\inst{1,2}\fnsep\thanks{\email{matteo.migliorini@pd.infn.it}} \and
    \firstname{Jacopo} \lastname{Pazzini}\inst{1,2,3,4} \and
    \firstname{Andrea} \lastname{Triossi}\inst{1,2} \and
    \firstname{Marco}  \lastname{Zanetti}\inst{1,2}
    \and
    \firstname{Alberto}  \lastname{Zucchetta}\inst{2}
}

\institute{
    Department of Physics and Astronomy ``Galileo Galilei'', Padova University, Italy
\and
    National Institute for Nuclear Physics, Padova Division, Italy
\and
    Department of Industrial Engineering, Padova University, Italy
\and
    Department of Information Engineering, Padova University, Italy
}

\abstract{%
   {%\footnotesize
    This work describes an online processing pipeline designed to identify anomalies in a continuous stream of data collected without external triggers from a particle detector. The processing pipeline begins with a local reconstruction algorithm, employing neural networks on an FPGA as its first stage. Subsequent data preparation and anomaly detection stages are accelerated using GPGPUs.
    As a practical demonstration of anomaly detection, we have developed a data quality monitoring application using a cosmic muon detector. Its primary objective is to detect deviations from the expected operational conditions of the detector. This serves as a proof-of-concept for a system that can be adapted for use in large particle physics experiments, enabling anomaly detection on datasets with reduced bias.

    %%OLD
    %This study presents a comprehensive computational framework designed for real-time analysis of high-throughput data streams developed using a small-scale drift tube detector setup. The work addresses the challenges of data volume and computational efficiency by employing a two-tiered data processing structure. Local reconstruction algorithms, executed on FPGAs, serve as the first stage, while subsequent data preparation for anomaly detection is accelerated using GPGPUs. The underlying goal is running the analysis algorithm, the New Physics Learning Machine, to detect deviations from the expected data distribution using the unfiltered $40\,\si{\mega\hertz}$ data stream.
    %The anomaly detection algorithm is used to monitor data quality, evaluating the integrity and performance of the detector. The algorithm shows sensitivity to induced malfunctions in the detector system, correlating with the severity of these malfunctions. Performance metrics further indicate that the system can maintain real-time data processing speeds while offering promising scalability for larger setups.
    %The developed framework serves as a proof-of-concept for a scalable, efficient system adaptable to larger particle physics experiments. It enables using less biased datasets for analysis, thereby offering a systematic approach to advancing experimental capabilities in exploring unexamined sectors in high-energy physics.
   }
}
\maketitle

% ---------------- Introduction

\section{Introduction}
\label{intro}

The sensitivity of modern high-energy physics experiments to New Physics is often limited by the hardware-level triggers used to select data online, resulting in a bias in the data collected. However, an online filtering stage is commonly needed to reduce the enormous throughput of complex data the detectors produce to make it manageable from a storage and offline computing perspective. Therefore, the deployment of efficient data acquisition systems integrated with online processing pipelines is instrumental in increasing the experiments' sensitivity to the discovery of any anomaly or possible signal of New Physics. In designing such systems, combining heterogeneous processing elements, including Field Programmable Gate Arrays (FPGAs) and General Purpose Graphics Processing Units (GPGPUs) \cite{HeterogeneusDagostinoCesini}, is key to sustaining the large throughput of unfiltered raw data.\\%
In this work, we present the first implementation of an end-to-end infrastructure that continuously acquires data from an experimental setup and processes it online, looking for statistical anomalies using machine learning. The goal is to develop and test a pipeline performing, as a first example, data quality monitoring (DQM). However, the final target of this project is to use this system to perform anomaly detection for new physics searches on large particle physics experiments. The infrastructure described in this paper is deployed at the INFN Legnaro National Laboratory (LNL). It reads out data from a reduced-sized version of the drift tube muon detector of the CMS experiment at CERN \cite{CmsExperiment, lemma}. An FPGA is in charge of collecting the data stream, clustering signals associated with the passage of a muon through the detector and producing candidate stubs \cite{MuonTriggerFpga}. Candidate events are then reconstructed, and all muon hits, along with muon stubs, are analyzed online by an algorithm deployed on a GPU to perform unbiased data exploration and statistical anomaly detection. The New Physics Learning Machine (NPLM) \cite{DAgnolo:2018cun,DAgnolo:2019vbw} technique is used to evaluate the compatibility between incoming batches of experimental data and a reference sample representing the expected behavior of the data under standard detector conditions. In the specific case of the LNL test stand, the NPLM algorithm uses as a reference sample a dataset gathered in normal detector conditions; deviations from the normal behavior, if detected, are characterized and statistically mapped to known sources of detector malfunctioning within a given degree of confidence. Unexpected behaviors that might signal the presence of new anomalies can be singled out if the observed discrepancy doesn't match any of the expected detector malfunctions.\\%
%{\color{red} MM: Aggiungere qualcosa/link migliore con prossimi capitoli?}
%The system is currently dealing with the limited throughput originated by the cosmic muon flux; nevertheless, all components of the readout chain are designed to scale up and eventually be employed in experiments at the LHC.
%In this contribution, we describe the technical implementation of the online processing pipeline and assess the performance of its most critical components.

% ---------------- Experimental setup

\section{Experimental setup}
\label{experimentalSetup}

The pipeline described in this work reads and processes the digitized signals produced by a muon telescope composed of a set of drift tube (DT) detectors. The DTs used in this work were built at the Legnaro National INFN Laboratories and inspired by those used in the CMS experiment, with which they share the same underlying design and configuration.
These detectors are designed to provide a small footprint ( roughly $70 \times 70\,\si{\cm}^2$) muon tracking system. They can be deployed in a number of different configurations, such as a muon spectrometer in conjunction with a magnetic field, or as a telescope for cosmic muons. Each DT chamber comprises 4 layers of 16 cells, totaling 64 cells per chamber. The signals produced by each cell (referred to as \textit{hit}) are amplified, discriminated, and shaped according to the LVDS standard. Two Xilinx VC707 evaluation boards are used to implement the time-to-digital conversion (TDC) of the hits, where each VC707 receives signals from 128 DT channels. The data stream of each VC707 board is serialized with the GBTx-FPGA protocol \cite{Baron:2009vbr} and transmitted via optical links to a Xilinx KCU1500 evaluation board mounted on the PCI express Gen-4 bus of a server where the processing and anomaly detection application is run. The server is a Dell PowerEdge R750 equipped with 2xIntel Xeon Gold 5318Y CPU @ 2.10GHz and 16$\times$16GB DIMM DDR4. 

% ---------------- Processing

\section{Reconstruction and data preparation}
\label{reconstruction}

Hits are received by the KCU1500 and deserialized before being processed in two stages. First, the local reconstruction of the muon segment is performed on the back-end evaluation board using a neural network-based algorithm. Hits and segments are merged into a single data stream and transferred to the server memory through Direct Memory Access (DMA) to avoid burdening the server CPU. 
In the second stage, a GPGPU is used to perform the data preparation for the anomaly detection application. 

\subsection{Muon reconstruction on FPGA}

The first reconstruction stage aims at identifying the time of passage of the muon, as well as its local position and crossing angle. Leveraging on the constant drift velocity inside each cell and the staggered geometry of the layers, the generalized mean-timer technique \cite{meantimer} can be used to find the absolute time of passage of a particle by combining the time information of $3$ or more hits. However, a left-right ambiguity with respect to the wire position is present and needs to be solved in order to solve the equations and find the track parameters. 
% OLD
%This can be used to translate the hit time into a spatial position inside the cell. However, a left-right ambiguity with respect to the wire position is still present and needs to be solved in order to extract the slope and intercept of the track. \\%
% daje
\begin{figure}
    \centering
    \includegraphics[width=0.9\textwidth]{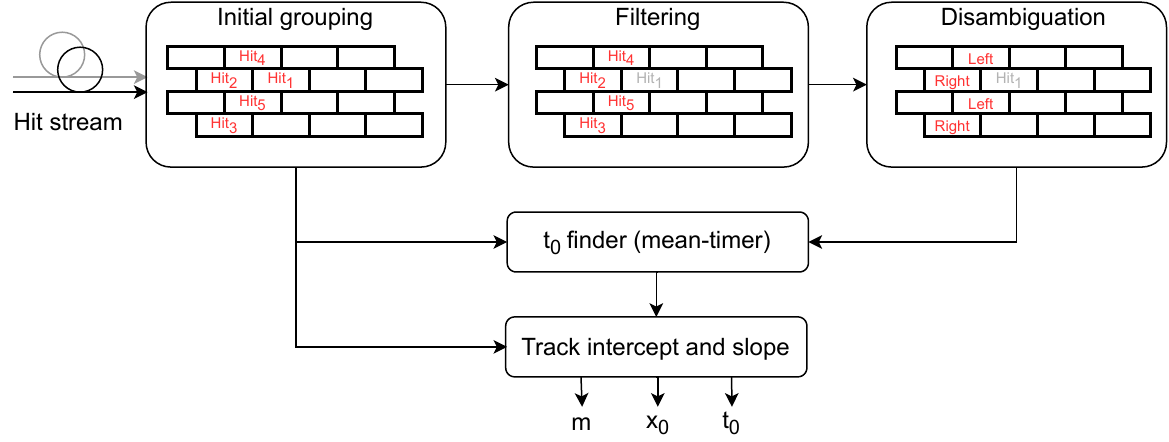}
    \caption{\small Schematic representation of the algorithm for each macro-cell. In the initial grouping, hits are collected from the stream and positioned in the macro-cell. This is then fed to filtering and disambiguation blocks where noise is rejected, and the left-right ambiguity is solved. This information is used to compute the muon crossing time and track parameters.}\vspace{-0.5cm}
    \label{fig:fpga-reco}
\end{figure}
The reconstruction algorithm implements a ``hybrid method'': two neural networks are used to filter hits and assign the correct laterality pattern starting from the hits time and channel information. Once the correct hit and laterality combination is found, the mean-timer equation univocally associated with that specific pattern is used to find the track parameters. Without the neural networks, all the possible pairs combination-equation need to be probed, resulting in a large combinatorial. A diagram of the algorithm is shown in Figure \ref{fig:fpga-reco}.
The first module of the algorithm, called \textit{initial grouping}, collects hits from the continuous stream and organizes them in time-coherent $4\times 4$ macro-cells, a set of neighboring cells fully containing all the possible patterns of a muon. Each macro-cell is then fed to the neural network blocks. The first, \textit{filtering}, retains only the hits produced by the passage of a muon. The hits passing the filtering step are then processed by the disambiguation network, where the correct laterality is predicted. From here, the crossing time $t_0$, intercept $x_0$, and slope $m$ are computed.\\%
To reduce the FPGA resource utilization, models were trained using \textsc{QKeras} \cite{Qkeras}\cite{sft_qkeras} and iteratively pruned during the training phase. Finally, the package HLS4ML \cite{HLS4ML}\cite{fastml_hls4ml} was used to produce the High-Level Synthesis code for the models. The inference time of the two neural networks is 2 clock cycles at $40\,\si{\mega\hertz}$ each, i.e., $50\,\si{\ns}$. However, the system is a pipeline with an initiation interval of 1, meaning that a new input can be accepted every clock cycle. Moreover, the aim of this stage is to enrich the stream of hits with the reconstructed stubs and not to filter as a trigger would do. 

\subsection{Data preparation on GPGPUs using Rapids}

\begin{figure}
    \centering
    \sidecaption
    \includegraphics[width=0.6\textwidth]{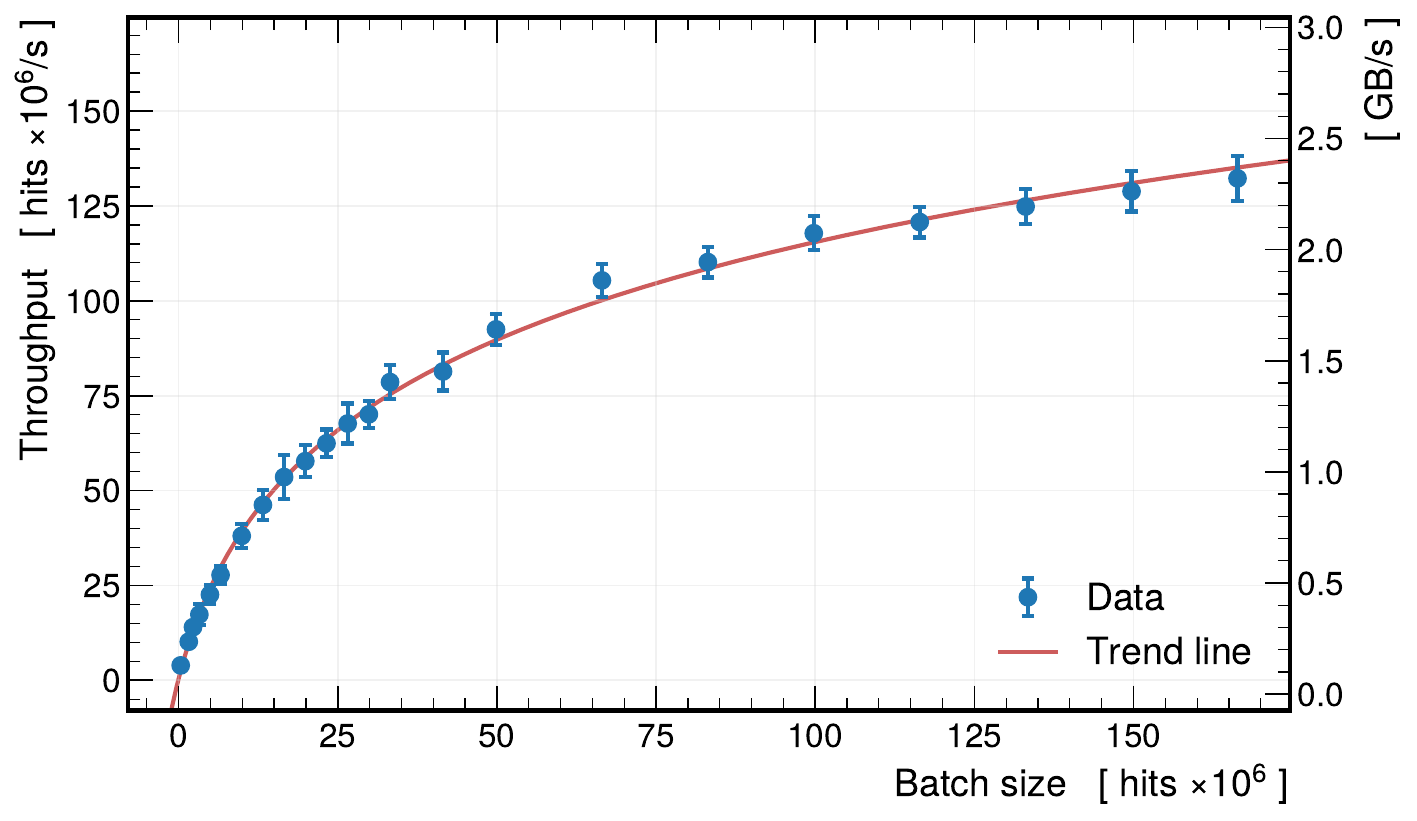}
    \caption{Throughput performance of the preprocessing algorithm in relation to batch sizes, measured in millions of hits. Throughput is quantified in terms of processed hits (left $y$-axis) and gigabytes (right $y$-axis) per second of processing time.}\vspace{-0.2cm}
    \label{fig:prepro-throughput}     
\end{figure}

After the reconstruction, all the hits and stubs are transferred to the server memory via DMA. % su RAMDisk?
Before being fed to the NPLM algorithm, they need to be aggregated and reformatted. This pre-processing consists mainly of identifying events, e.g., time slices containing muon stubs, filtering spurious hits, and computing new higher-level quantities describing the event. This can be performed leveraging distributed computing frameworks using dataframe-like data structures and operations, as it has been described in \cite{2022NIMPA103666869M}. This task can be further accelerated using GPGPUs thanks to \textsc{CUDA} dataframe (\textsc{cuDF}) library implementation in NVIDIA \textsc{Rapids} \cite{Rapids}.
The \textsc{cuDF} library, built on top of Apache Arrow columnar memory format, is used to perform dataframe operations on a GPGPU. %It provides C++ and Python libraries, whereas the latter provides a more user-friendly interface at the price of some limitations.
Along with dataframes operations, I/O for standard data formats such as Apache Parquet is implemented. Moreover, the basic set of functionalities can also be extended by writing custom kernels using either \textsc{CUDA} or \textsc{CuPy}/\textsc{Numba}.
The processing steps used in this work have been easily ported to \textsc{cuDF}, as they consist mainly of simple aggregations, filtering, and column-wise operations. The performance of this pipeline stage can be seen in Figure \ref{fig:prepro-throughput}. The time needed to read and process data is measured by changing the batch size, i.e., the number of hits in the dataset processed together. A throughput of over $2\,\si{\giga\byte/\second}$ is achieved when working with large batch sizes of the orders of millions of hits. Getting this amount of data from our test setup using cosmic rays would require running the acquisition for over 2 hour. However, for large experiments, the same amount of data can be collected in seconds. The goal of online processing is to reduce data, i.e., computing a higher level representation of the data, to keep the throughput manageable for the anomaly detection application.

%A large throughput is achieved since no computationally heavy operation is performed. This can be extended by running, for example, a tracking step, i.e., combining stubs from the different chambers. \\

% Moreover, one possible of this system is to transfer data from the FPGA directly to the GPU memory, skipping the intermediate step of copying data to the ramdisk, at the price of a more complex system.
% Preprocessed data can now be ingested by the NPLM algorithm, which will be run in the same GPU.

% {\color{red} NL: Espandere?? Dettagli?? Forse manca qualche numerello magico di riferimento, come si diceva il preprocessing reduce da X GB/s a Y MB/s per fittare i requirements dell'analisi.}

% ----------------

\section{Anomaly detection}
\label{anomalyDetection}

%%%% OLD
% Implementing an online anomaly detection system on an unfiltered 40 MHz muon data stream presents a practical opportunity for data quality monitoring (DQM). At its core, DQM is similar to exploring New Physics. Anomalies in DQM indicate detector issues, while New Physics suggests unexpected phenomena. 

% $40\,\si{\mega\hertz}$
At large-scale collider experiments, utilizing an unfiltered muon data stream would enable the exploration of yet-unobserved anomalies, which are commonly filtered away by trigger systems. For the purpose of testing the full chain implementation on our small-scale experimental setup, we recast this advantage into a DQM framework, in which detecting anomalies serves as a real-time evaluation metric for the detector's performance.

\subsection{Dataset and anomalies}\label{sec:dataset}

Following the track reconstruction process, data is transformed from a hit-centric representation to an event-oriented one. Events are thus defined by their four drift times—one from each layer—and the crossing angle of the muon with the vertical axis.
To mimic the most common sources of detector anomaly, we then induced potential malfunctions within the system. 
By altering the voltage of the cathodic strips and the front-end thresholds, we introduced malfunction scenarios at varying intensities: 75\%, 50\%, and 25\% of their typical operational values. 
The datasets procured under these anomalous conditions were treated with the same rigorous processing as those obtained under regular conditions. As part of our data acquisition campaign, we systematically gathered information across six distinct configurations, capturing approximately $10^4$ events for each setup. Additionally, to establish a robust baseline for comparison, we collected approximately $3 \times 10^5$ events from the detector's standard operational conditions.\footnote{Datasets available at \href{https://doi.org/10.5281/zenodo.7128223}{\texttt{https://doi.org/10.5281/zenodo.7128223}}.} 
The datasets are further discussed in Ref.~\cite{DQMpaper}.

\subsection{Anomaly detection methodology}

In data quality monitoring, one aims to assess the operational state of the detector, which has a direct impact on the quality of the data gathered. This evaluation can be formally framed as assessing the alignment of a dataset, denoted as $ \D $, with the expected statistical distribution under standard operational conditions, represented as $ P(x\,|\,\text{R}) $.
 The direct knowledge of the reference distribution $ P(x\,|\,\text{R}) $, however, is unavailable. We thus exploit a reference dataset $ \R $ that stands in for this distribution. Therefore, our method compares the datasets $ \D $ and $ \R $ to see if they originate from the same statistical distribution.

In our investigation, we employ a reference dataset of fixed size $\N_\R=2000$, extracted from a larger pool of approximately $ 3 \times 10^5 $ muon data collected under detector conditions labeled as \textit{standard}. We then monitor batches of varying sizes, $\N_\D=250$, $\N_\D=500$, and $\N_\D=1000$, for all the anomalous configurations reported in Section~\ref{sec:dataset}. 

\paragraph{The NPLM approach}

To compare a data batch $ \D $ with a reference sample $ \R $, we employ the NPLM algorithm \cite{DAgnolo:2018cun, DAgnolo:2019vbw}. This algorithm calculates a test statistic by quantifying the log-likelihood ratio between the reference hypothesis and the actual, yet unknown, distribution generating the observed data batches. It does so through a machine-learning model $ f_{\ww}(x) $, which acts as a universal function approximator parameterized by trainable parameters $ \ww $. It introduces alternative data distributions $P(x\,|\,\text{H}_{{\ww}})$, deviating from $P(x\,|\,\text{R})$ if $\D$ and $\R$ stem from different distributions.

% \begin{equation}\label{eq:log-loss}
%     l(y,\,f_{\ww}(x)) = 
%     (1-y)\,\left(1+\frac{\N_\D}{\N_\R}\right)\log\left(1+e^{+f_{\ww}(x)}\right) + 
%     y\,\left(1+\frac{\N_\R}{\N_\D}\right)\log\left(1+e^{-f_{\ww}(x)}\right)  \,.
% \end{equation}

% Here, $ y=0 $ designates data points in $ \R $, while $ y=1 $ is assigned to data in $ \D $. Moreover, the optimization involves incorporating L2 regularization, scaled by a weight factor $ \lambda $ termed as the regularization weight.

The model $ f_{\ww}(x) $ can be constructed using either neural network or kernel methods and trained using a weighted logistic loss function. Notably, the \textsc{Falkon} library \cite{falkonlibrary2020}\cite{sft_falkon} significantly enhances the efficiency of kernel methods. This library employs Gaussian kernels with a tunable width $ \sigma $. Furthermore, this library introduces several simplifications that enhance computation, especially for GPU-based training. Noteworthy among these is the Nystr\"om approximation, which reduces computational complexity by enabling solutions composed of weighted combinations of selected data points. These points, termed Nystr\"om centers, are randomly chosen from input data. The selection of the number of centers, $ M $ ($ M \le \N $), is a hyperparameter demanding careful consideration. For further insights, refer to Ref.~\cite{falkonlibrary2020}. 

\paragraph{Hyperparameter tuning}

For hyperparameter selection, we rely exclusively on data from the reference working condition to ensure model independence. The tuning process involves the optimization of the following:

$\,\,1.\,\,$ The number of centers, $M$, which balances model expressiveness and training speed, by seeking the largest $M$ that improves expressiveness without excessive training time. 

$\,\,2.\,\,$ The Gaussian width $\sigma$, determined as the 90th percentile of pairwise distances among reference-distributed data points. 

$\,\,3.\,\,$ The regularization parameter $\lambda$, minimized while maintaining training stability. 

\paragraph{Calibration and p-value extraction}

After hyperparameter optimization, the model undergoes training to adapt its parameters $ \ww $ to approximate observed data. This yields trained parameters $ \Hat{\ww} $ characterizing the best-fit hypothesis $ \text{H}_{\Hat{\ww}} $. The test statistic associated with the log-likelihood ratio for a data batch $ \D $ and reference sample $ \R $ is then computed as:

\begin{equation}\label{eq:test-statistic}
    t(\D)=2\sum_{x\in\D}\log\frac{P(x\,|\,\text{H}_{\Hat{\ww}})}{P(x\,|\,\text{R})}=2\sum_{x\in\D}f_{\Hat{\ww}}(x)\,.
\end{equation}

The numerical value of $t(D)$ has no inherent meaning. To extract physical insights, it has to be calibrated and referenced to the test statistic distribution under the reference hypothesis, $ P(t\,|\,\text{R}) $.
The test statistic calibration proceeds by considering larger positive values of $ t $ as more likely if the best-fit alternative distribution $ P(x\,|\,\text{H}_{\Hat{\ww}}) $ fits the data better than the reference distribution $ P(x\,|\,\text{R}) $. This suggests that $ \D $ might not adhere to $ P(x\,|\,\text{R}) $. To estimate $ P(t\,|\,\text{R}) $ empirically, we generate artificial data batches (termed \textit{toy} datasets) of the same size $ \N_\D $ as real data batches, train the model, and compute $ t $ for each. By creating histograms of $ t $ values from toy datasets, we estimate $ P(t\,|\,\text{R}) $. Furthermore, considering $ P(t\,|\,\text{R}) $ approximated by a $ \chi^2 $ distribution, we fit the empirical distribution of $ t $ values to obtain a continuous test statistic distribution under reference conditions.

To quantify the discrepancy between $ \R $ and $ \D $, we calculate the $ p $-value:

\begin{equation}\label{eq:pvalue}
    p\,[\,t(\D)\,] = \int_{t(\D)}^{\infty}\, P(t'\,|\,\text{R})\,\text{d}t'\,.
\end{equation}

The $ p $-value measures the likelihood that a reference-distributed batch yields a test statistic value more improbable (i.e., larger) than the observed $ t(\D) $.

\subsection{Results and Performance}
\begin{figure}
    \centering
    \sidecaption
    \includegraphics[width=0.6\textwidth]{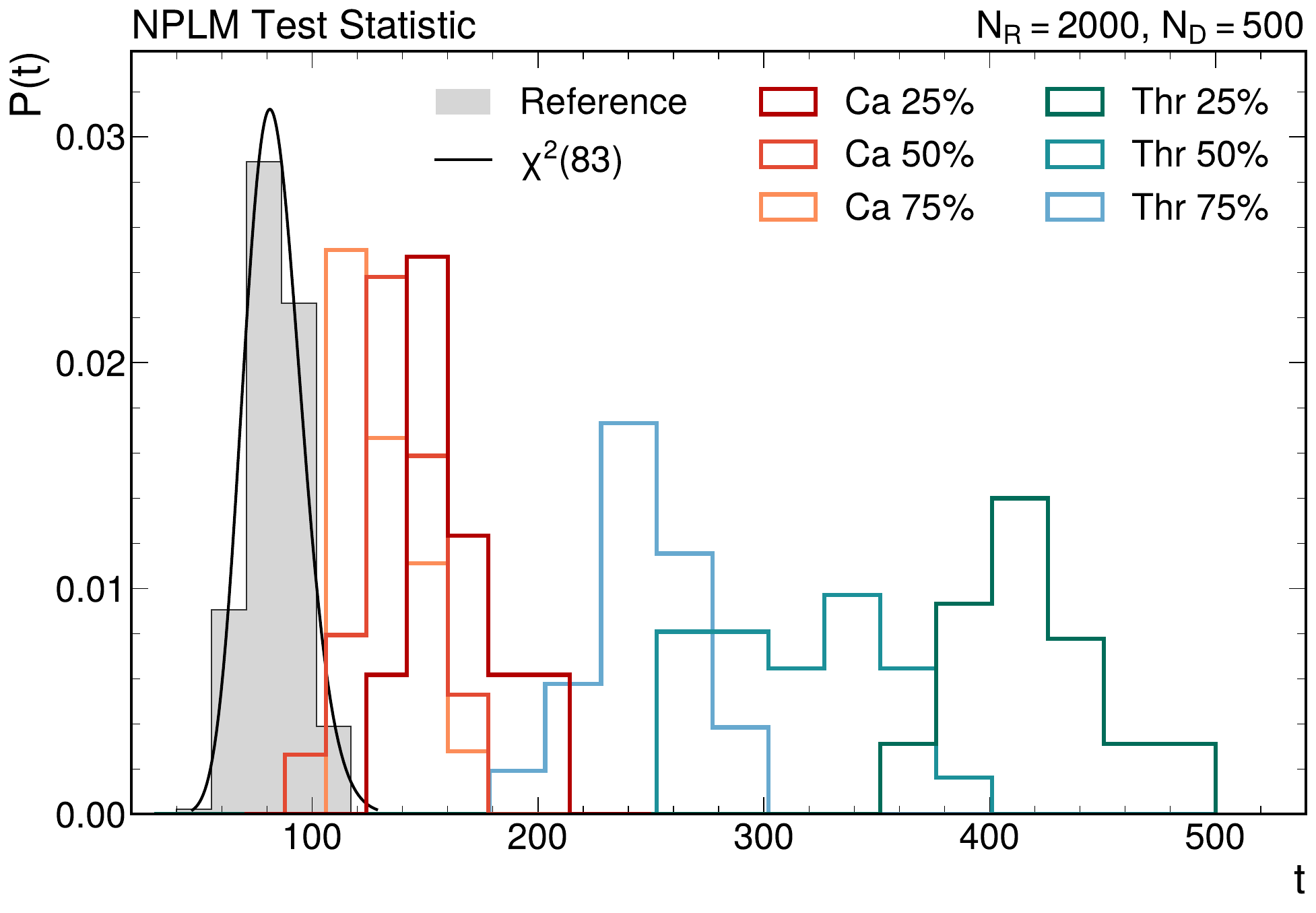}
    \caption{Distribution of the test statistic for a data batch size of $\N_\D=500$. Grey histograms represent the empirical test statistic distribution $P(t\,|\,\text{R})$. Histograms in shades of red and green are test statistics of anomalous batches corresponding to the cathodes and thresholds anomaly classes, respectively.}\vspace{-0.2cm}
    \label{fig:tresults}     
\end{figure}

\paragraph{Anomaly detection performance}

After calibrating the test statistic in Eq.~\ref{eq:test-statistic} using toy datasets from the standard detector conditions, we sample (without replacement) a few datasets from each anomaly category and fill a histogram of the output $t(\D)$ values. Then, we compute the $p$-value of the median of the test statistic distribution using Eq.~\ref{eq:pvalue} and a fitted $\chi^2$ distribution approximating $P(t\,|\,\text{R})$. Figure~\ref{fig:tresults} shows the NPLM test statistic distribution for the cathodes and thresholds anomalies for the batch size configuration of $\N_\D=500$. Further details are available in Ref.~\cite{DQMpaper}.

From these histograms, we observe that the test statistics for anomalous batches, highlighted in Fig.~\ref{fig:tresults} in shades of red and green for the cathodes and thresholds anomaly classes, respectively, are noticeably distinct from the reference distribution. This distinction validates the monitoring algorithm's capability to identify anomalies. 
To provide a quantitative measure, we assess the median values of the \( t(\D) \) distributions corresponding to the 75\%, 50\%, and 25\% anomaly configurations. We find that as the severity increases from 75\% to 50\% to 25\%, the median of the test statistic distribution becomes progressively larger, increasingly diverging from the reference \( t \) distribution $P(t\,|\,\text{R})$. This observed trend quantitatively confirms the algorithm's enhanced sensitivity to anomalies with escalating severity of detector failure.

%% OLD
%Furthermore, it is evident that as the severity of the detector failure amplifies, so does the sensitivity to anomalies. This is visually confirmed by comparing the $t(\D)$ distributions corresponding to the 25\% and 75\% configurations. The former consistently displays larger test statistic values, diverging further from the reference $t$ distribution, while the latter remains relatively closer to $P(t\,|\,\text{R})$.

\paragraph{Training time performance and scalability}

\begin{figure}[!t]
    \begin{subfigure}[c]{0.5\textwidth}
        \centering 
        \includegraphics[width=0.97\textwidth]{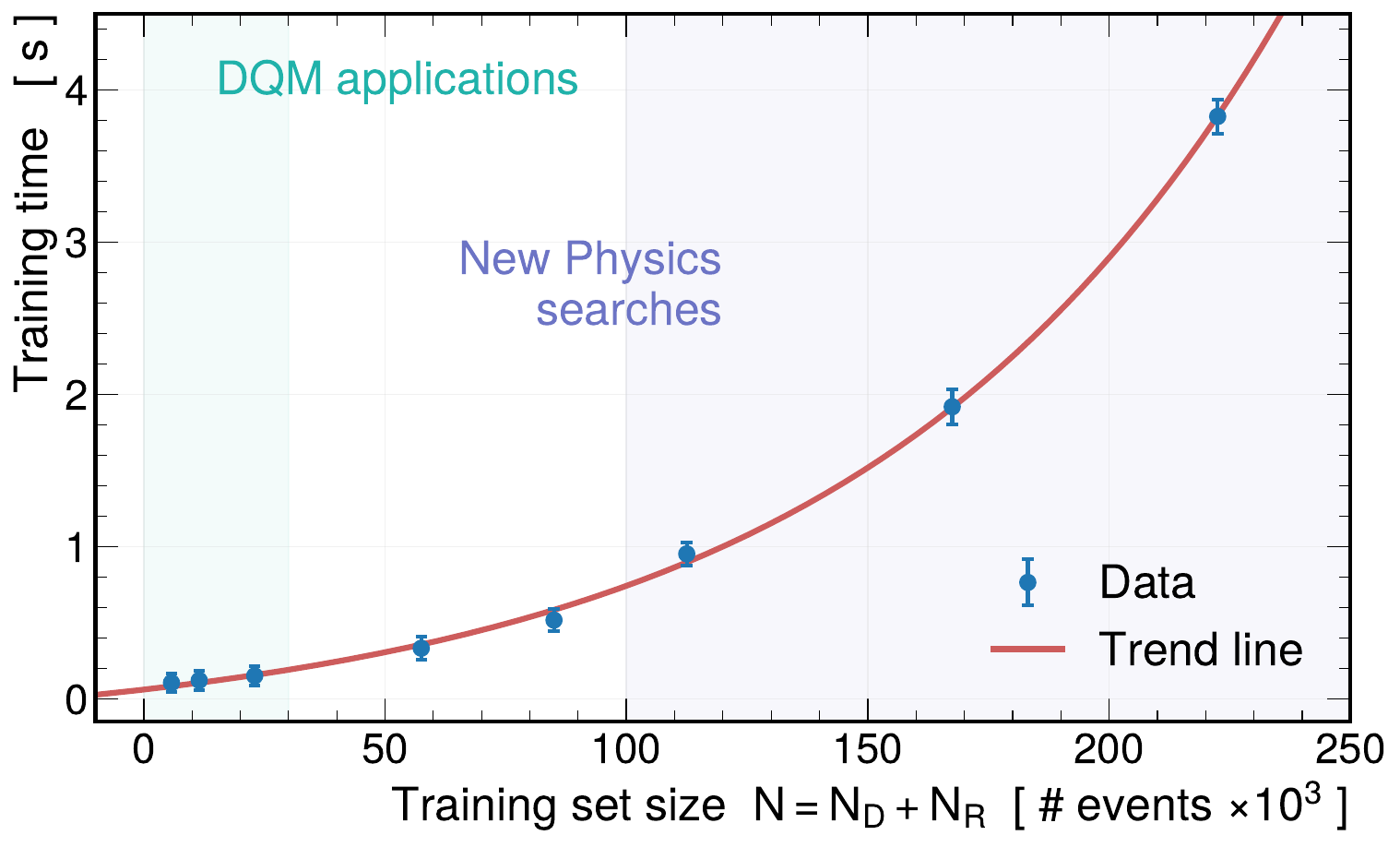}
        \caption{}
        \label{fig:perftimesmall}
    \end{subfigure}%
    \begin{subfigure}[c]{0.5\textwidth}
        \centering 
        \includegraphics[width=0.97\textwidth]{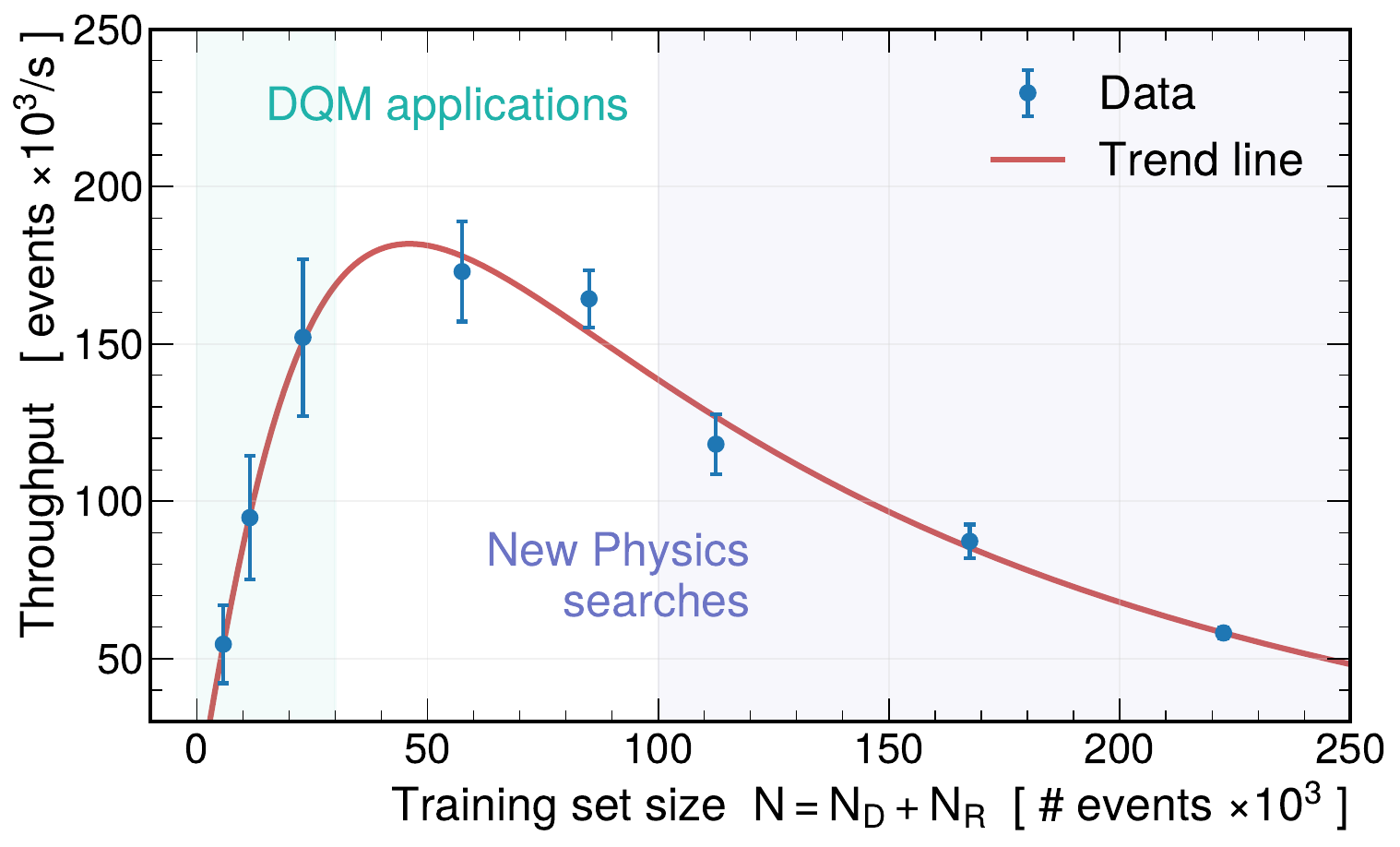}
        \caption{}
        \label{fig:perftrhoughputsmall}
    \end{subfigure}%
    \caption{\small Left (a): Training time as a function of the training set size $ \N $ for the NPLM algorithm.
    Right (b): Throughput, calculated as the number of muon events divided by the average training time, plotted against the training set size $ \N $. Shaded areas represent regions typical for DQM applications and New Physics searches using NPLM.}\vspace{-0.2cm}
    \label{fig:perf}
\end{figure}

% We now examine the time performance relative to the training set size, denoted $ \N = \N_\D + \N_\R $. This training set size prominently influences the operational dynamics of the non-parametric kernel-based NPLM. 

%% OLD
% To assess the scalability of training execution time for larger experiments with significantly higher throughput, resulting in larger data batches to analyze with fixed latency, we systematically adjusted $ \N_\R $ and $ \N_\D $.

As the natural objective of this work is to extend its deployment to large-scale experiments with significantly higher throughput, it becomes crucial to evaluate the scalability of training execution time. Accordingly, we systematically adjusted $ \N_\R $ and $ \N_\D $ to assess the system's performance in handling larger data batches within a fixed latency.
We increased the total event count $ \N = \N_\D + \N_\R$ in the training set, primarily by increasing $ \N_\R $, while maintaining a relatively stable ratio $ \N_\D\,/\,\N_\R $. Subsequently, we tuned the model hyperparameters for each unique configuration, spanning $ \N $ from approximately $ 2 \times 10^3 $ to $ 230 \times 10^3 $, employing the same hyperparameter selection procedure.
We ran the NPLM algorithm on distinct training sets $\OO(1000)$ times for each configuration. We use the average training time as our primary metric. The standard deviation provides additional insights into the variability amongst repeated independent training. 
This approach provided the training time trend relative to the training set size $ \N $, as shown in Figure~\ref{fig:perftimesmall}. Although the training time follows an exponential trend with respect to $\N$, the absolute values of the training times for the range of batch sizes examined are sufficiently low. This supports the algorithm's suitability for real-time applications.
Additionally, we determined the throughput by dividing the number of analyzed events by the average training time. The throughput as a function of \( \N \) is illustrated in Figure~\ref{fig:perftrhoughputsmall}. We observe that the throughput initially increases for smaller batch sizes, peaks in the range of $50\,000$ to $100\,000$ events, and subsequently decreases. Note that the derived metrics for training time and throughput serve as benchmarks that are inherently tied to the particular system configuration and computing infrastructure used in our study.

% OLD
%Additionally, we calculated the throughput by dividing the number of events analyzed by the average training time. The plotted throughput as a function of $ \N $ is presented in Figure~\ref{fig:perftrhoughputsmall}.

% NL: A STO PUNTO DROPPEREI PER MANCANZA DI SPAZIO
%Two specific ranges are shaded within the plots: a light green region corresponds to typical $ \N $ values in data quality monitoring applications. In contrast, a light violet area highlights values commonly associated with New Physics searches using NPLM. This visual differentiation effectively underscores the disparities between these regions concerning both training time and throughput. 

% ---------------- Conclusions
\section{Concluding remarks}
\label{conclusions}

The primary objective of this study was the implementation of an online data processing pipeline for anomaly detection on a trigger-less data stream. As a first example, data generated by a muon detector is considered. The pipeline architecture comprises a two-stage data processing flow: an initial local reconstruction executed on an FPGA, followed by data preparation and anomaly detection stages, both benefiting from GPGPU acceleration.
We leveraged the New Physics Learning Machine (NPLM) algorithm to create a data quality monitoring application that identifies induced detector malfunctions spanning various severity levels. 
The effective implementation of the NPLM algorithm for data quality monitoring using unfiltered, high-throughput data streams suggests its applicability to broader research objectives. Specifically, the same computational pipeline can be adapted to search for New Physics phenomena in large-scale, high-energy physics experiments.
% The successful application of NPLM in this context underscores its potential for broader research objectives, particularly in the pursuit of uncovering New Physics phenomena. The same principles facilitating precise data quality monitoring can be harnessed for detecting anomalies within unfiltered, high-throughput data streams, thereby extending the versatility of this system to large-scale, high-energy physics experiments.
To enhance the pipeline's efficiency, several paths for improvement are available. 
Most notably, future works will focus on streamlining the data flow from the FPGA to the GPU memory by applying zero-copy data transfer models.

\section*{Acknowledgements}

{This research was supported by grants from NVIDIA and utilized NVIDIA A100 GPU. \\
This work is partially supported by ICSC–Centro Nazionale di Ricerca in High Performance Computing, Big Data and Quantum Computing, funded by European Union – NextGenerationEU.
}

\bibliography{bibliography}

\end{document}